\documentclass[12pt,preprint]{aastex}
\usepackage{epsfig,lscape}
\shorttitle{STELLAR POLLUTION}
\shortauthors{LI, LIN, \& LIU}
\begin{document}
 
\title{Extent of pollution in planet-bearing stars}
\author{S.-L. Li$^{1,2}$, D.~N.~C. Lin$^{2,3}$, and X.-W. Liu$^{1,3}$}
\affil{$^1$Department of Astronomy, Peking University, Beijing 100871, China}
\affil{$^2$Department of Astronomy and Astrophysics,
University of California, Santa Cruz, CA 95064, USA}
\affil{$^3$Kavli Institute for Astronomy
and Astrophysics, Peking University, Beijing 100871, China}
 
\begin{abstract}

Search for planets around main-sequence stars more massive than the Sun is
hindered by their hot and rapidly spinning atmospheres. This obstacle has been
sidestepped by radial-velocity surveys of those stars on their
post-main-sequence evolutionary track (G sub-giant and giant stars).
Preliminary observational findings suggest a deficiency of short-period hot
Jupiters around the observed post main-sequence stars, although the total
fraction of them with known planets appears to increase with their mass. Here
we consider the possibility that some very close-in gas giants or a population
of rocky planets may have either undergone orbital decay or been engulfed by
the expanding envelope of their intermediate-mass host stars. If such events
occur during or shortly after those stars' main sequence evolution when their
convection zone remains relatively shallow, their surface metallicity can be
significantly enhanced by the consumption of one or more gas giants. We show
that stars with enriched veneer and lower-metallicity interior follow slightly
modified evolution tracks as those with the same high surface and interior
metallicity. As an example, we consider HD\,149026, a marginal post-main
sequence 1.3-$M_\odot$ star. We suggest that its observed high (nearly twice
solar) metallicity may be confined to the surface layer as a consequence of
pollution by the accretion of either a planet similar to its known
2.7-day-period Saturn-mass planet, which has a 70-$M_\oplus$ compact core, or a
population of smaller mass planets with a comparable total amount of heavy
elements. It is shown that an enhancement in surface metallicity leads to a
reduction in effective temperature, in increase in radius and a net decrease in
luminosity. The effects of such an enhancement are not negligible in the
determinations of the planet's radius based on the transit light curves.
Finally, we show that the extent of pollution can be inferred directly from
high precision distant determinations and transit observations.

\end{abstract}


\section{Introduction}

In the radial-velocity search for planets around nearby main sequence FGK dwarf
stars, it is well established that the detection probability of Jupiter mass
gas giant planets increases rapidly with the metallicity of the target stars
\citep{Gonzalez1997, Fischer2003, Santos2003}. There are two scenarios for this
observational correlation: 1) the host stars of planetary systems are polluted
by infalling planets \citep{Sandquist02}, 2) gas giant planets are
preferentially formed around metal-rich stars \citep{IdaLin2005}.

For the first (pollution) scenario, there are many potential avenues which may
lead to planetesimal or protoplanet accretion by their host stars. For example,
during the epoch of protoplanet formation, planetesimals and protoplanets
undergo orbital decay as they interact with their nascent disks
\citep{Goldreich1980, Ward1984, Lin1986}. Unless the migration of those
entities can be stalled by their interaction with their host stars
\citep{Lin1996}, they have a tendency to be accreted \citep{Ida2004}. Since T
Tauri stars have deep convective regions, any accreted metal-rich protoplanets
during the first few Myr formation time of the star are dissolved in the
envelope and the excess metals are mixed by convection and homogenized
throughout the envelope.

On the time scale of $\sim 30$Myr, the convection zone of a solar-type star is
reduced to contain only the top 2\% of the total stellar mass and $\sim 100
M_\oplus$ heavy elements \citep{Ford1999}. This time scale is shorter and the
total mass of the convection zone is smaller for F than for G dwarf stars. In
systems which contain eccentric gas giants, the disk depletion process also
leads to a sweeping secular resonance \citep{Ward1976, Heppenheimer80,
Ward1981, Nagasawa2000, Nagasawa2001} which can cause the residual
planetesimals to migrate towards their host stars \citep{Nagasawa05, Zhou05}
and excite the gas giants' eccentricities \citep{Nagasawalin2005}. After the
gas depletion, long term dynamical instability can also lead to orbit crossing
and close encounters which can scatter the residual planets into their host
star (Chambers 2001). Finally, most stars form in clusters which are eventually
disrupted under Galactic tides. Stellar encounters can also lead to
perturbations which can destabilize planetary systems and induce residual
planetesimals bombardment onto their host stars \citep{AdamsL03}. When those
planets and planetesimals enter the atmosphere of their host stars, they become
disintegrated as a consequent of Rayleigh-Taylor and shearing instabilities
\citep{Sandquist02}. The metal-enriched planetary debris rapidly dissolves and
becomes mixed with the gas in the convective envelope near the stellar surface.
In G dwarfs, this process can significantly enhance the observable metallicity
on their surface.

For the second (threshold-core-accretion) scenario, gas giant planets form more
prolifically around metal-rich stars. Presumably all the contents of those
stars were accreted via proto-stellar disks. Heavy elements first condense into
grains which not only evolve through cohesive and disruptive collisions with
each other but also undergo orbital decay as a consequence of gas drag. The
efficiency of grain-planetesimals conversion increases with the metallicity of
the disk because both collision rate and orbital decay time scale also increase
with the disk metallicity \citep{Supulver00}. Gravitational instability in dust
layers is also more likely to occur in metal rich disks \citep{Sekiya1983,
Youdin2002}. However, the threshold metallicity for the onset of gravitational
instability may be considerably larger than those of the most metal-rich stars
\citep{Garaud2004}. The additional planet-building material also enhances the
growth rate of planetesimals and increases the asymptotic isolation mass of
proto-planetary embryos around metal-rich stars and therefore promotes the
emergence of gas giant planets \citep{Ida2005, Ida2007}.

A direct method to distinguish between these two scenarios and to
quantitatively determine the extent of heavy element pollution is to
observationally deduce the metallicity dispersion among similar mass stars
within a given open cluster. The observed chemical homogeneity among solar-type
stars in the Pleiades and IC 4665 clusters \citep{Wilden2002, Shen2005} places
a stringent upper limit ($< 10 M_{\oplus }$) on the added heavy elements that
they might have acquired after they enter the main sequence evolutionary track.
In addition, the formation of planets requires proto-stellar and debris disks
to retain at least comparable amount of heavy elements ($\sim 100 M_{\oplus }$)
as that contained in the solar system planets \citep{Ida2004}. The observed
chemical homogeneity among the cluster stars challenges this requirement.
However, despite several attempts to search for planets in the Hyades
\citep{Guenther2005}, only one planet has been found around one star
\citep{Sato2007}. Therefore, the stringent upper limit in the metallicity
dispersion does not provide sufficient information to place an upper limit on
the amount of pollution onto the cluster stars.

A more subtle differential comparison is to search for any correlation between
the metallicity and mass of planet-bearing field stars. Since the mass of the
convective envelope of solar-type stars decreases with stellar mass, extensive
pollution should lead to a correlation between the spectral type and the
surface metallicity \citep{laughlin97}. Among the known GK dwarf stars
harboring planets, the lack of any spectral class-metallicity correlation has
been cited as evidence that the late accretion of planets and planetesimals
cannot account for the preferential association of planets with metal-rich
stars \citep{Fischer05}. Quantitatively, it limits the amount of late
planetesimal/planet accretion to be less than the total heavy elements
contained within the surface convection zone of planet-bearing stars. For
planet-bearing main-sequence host stars (mostly G and K dwarfs), the mass of
their convective envelope is $\ga 10^{-2} M_\odot$. So to yield any observable
variations in their surface metallicity, pollution of well over $100
M_{\oplus}$ metals are needed after the stars have begun their main sequence
evolution.

Although, the elevated metallicity of planet-bearing solar type stars can be
easily accounted for by the threshold-core-accretion scenario, pollution at
some level is unavoidable unless the planet formation process is highly
efficient in retaining all the planet building blocks. In the outer solar
system, the long-term perturbation of Jupiter and Saturn has resulted in the
outward scattering of most residual planetesimals and the formation of the
Kuiper-belt objects and Oort Clouds. Nevertheless, at least a fraction of this
population (a few $M_\oplus$) is also scattered inward to form the terrestrial
planets and to be accreted by the Sun. Due to observational selection effects,
most of the known extra-solar planets have periods less than those of the gas
giants in the solar system. In those relatively compact planetary systems, the
inward-scattered fraction is likely to be larger than that in the solar system
because close planetary encounters are less able to provide an adequate
dynamical boost to the planetesimals' orbital energy for them to overcome the
steeper stellar gravitational potential. The determination of this fraction is
particularly important in the calibration of the efficiency of planet-formation
process and the assessment on the fraction of stars which contain gas giant
planets.

Another motivation to search for evidences of planet consumption is the
theoretical expectation of a larger-than-observed population of close-in gas
giant planets \citep{Ida2005, Ida2007}. Although type II migration can account
for the overall origin of the short-period planets, their survival requires
halting mechanisms such as planet-star interaction for those with close-in
orbits and disk depletion for those with intermediate (a week to a year)
periods. The period distribution of these planets requires the disruption of a
significant fraction of gas giants which migrated to the stellar proximity.
Around systems with known close-in planets, there is an elevated possibility of
several prior generation gas giants which have migrated to even closer to their
host stars. These ``missing'' planets may either have migrated directly into
their host pre-main-sequence host stars or were halted initially and then
resumed their orbital decay on a much longer tidal evolution time scale. The
latter scenario would imply a general decline in the fraction of stars with
close-in gas giants as the systems age. Again, establishing evidence for gas
giant planet consumption is important in providing valuable clues on the
dynamical evolution of planetary systems.

Ideally, constraints on the extent of protracted metallicity pollution is best
placed on the intermediate-mass (earlier than F8) planet-bearing stars. The
accretion of a Jupiter-like planet or a few $M_\oplus$ planetesimals during the
main sequence life span of these candidate stars would significantly enhance
the metallicity of their shallow convective envelopes. However, atmospheres of
these stars are hotter than those of the solar-type stars and they often have
rapid spins. The former effect prevents the formation of many useful
spectroscopic lines whereas the latter introduces considerable Doppler
broadening. Both effects introduce challenges to the radial velocity surveys
for planetary companions around high-mass stars. Consequently, there were very
few F stars in the original sample based on which the spectral
class-metallicity correlation was established.

More recently, searches for planets around massive stars have shifted to target
such stars after they have evolved onto their sub-giant and giant tracks and
become G-giant stars. During this post-main-sequence phase, the atmosphere of
these stars becomes cool with relatively slow spins. Preliminary surveys
indicate that the fraction of these relatively massive G giant stars with
planets may be higher than that around the solar-type G dwarf main sequence
stars \citep{Sato2007, Johnson2007, Dollinger07}. In addition, all of these
planet-bearing stars have metallicity comparable to that of the Sun. Based on
the lack of any obvious metallicity-dependence in the frequency of
planet-bearing G giant stars, in contrast to the well-known metallicity-planet
frequency correlation amount the G-dwarf stars, \citet{Pasquini2007} suggest
that the planetary consumption may be a major cause for this dichotomy. In this
scenario, 1) the high surface metallicity among the main sequence stars is
mainly due to planet consumption and 2) the surface metallicity of
planet-bearing stars decreases during their post main sequence evolution as
their convective envelopes steepen and become more diluted.

There are some supporting evidences for planet consumptions among the G giant
stars. There is an apparent lack of short-period gas giant planets around these
stars despite the fact that they are much easier to be detected. In contrast,
several-dozen known planets around main-sequence stars have periods less than 5
days, with some as short as 1 day. The post-main-sequence expansion of their
host stars' envelope, the star-planet tidal effect may promote the orbital
decay of close-in gas giants by the excitation of both Hough and inertial waves
\citep{Ogilvie2007}. If such a process can lead to the consumption of
short-period gas giants, the accreted planetary debris would be retained near
the surface layer above the radiative interior. Eventually, the stellar
envelope engulfs the orbits of the close-in planets, induces their orbital
decay and disruption \citep{Livio2002}. The mass contained in the convection
zone of the post-main-sequence stars increases with the expansion of their
photospheric radius. When these stars enter the sub- giant phase, there would
be sufficient mass in the convective envelope to homogenize and reduce their
surface metallicity to their interior solar values. Nevertheless, between these
two epochs, {\it i.e.} shortly after they have evolved off the main sequence
but before they have entered the sub-giant track, the prior consumption of very
close-in metal-rich gas giants or a population of rocky planets around
sufficiently massive stars may lead to noticeable metallicity enhancement near
their surface layers.

In this paper, we identify a planet-bearing star, HD\,149026, as a transitional
(between main-sequence turnoff and the subgiant branch) system. The host of
this system is a post main sequence 1.3-$M_\odot$ star with nearly twice solar
metallicity and it has a $P=2.7$day period Saturn-mass planet (HD\,149026b)
with a 70-earth-mass core. In a companion paper \citep{Li2007}, we propose that
the large core structure of this planet was acquired through planetary mergers
after it has migrated to the proximity of HD\,149026. In this scenario, some
debris material and residual planetesimals are expected to be scattered into
the host star by such a close-in planet. The migration of HD\,149026b may also
have induced the orbital decay of planets and planetesimals along its migration
path \citep{Zhou05}. Long term dynamical instability could also have led to
planet accretion onto the host star. Based on these anticipations, we identify
HD\,149026 to be an ideal star to study the effect of pollution.

The present-day convection zone of HD\,149026 is an order of magnitude less
massive than that of the Sun ($0.02$\,$M_\odot$). The consumption of a few
earth mass of residual planetesimals, embryos, or a metal-rich gas giant can
easily increase the metallicity of this thin layer from the solar value to the
observed one ([Fe/H] $\sim 0.36$). We briefly describe the method for computing
the effect of pollution in \S2. We present the impact of the stellar structure
due to planetary accretion in \S3 and the effect on the stellar evolution
during the main-sequence turn off of HD149026. We show that within the range of
the present observational uncertainties, the internal [Fe/H] of HD\,149026 may
be comparable to that of the Sun. We also show that a surface pollution of this
star may modify its post-main-sequence evolutionary track and its present-day
mass-radius relation. Based on these stellar models, we outline a set of
self-consistent observational tests which may be used to check the merger and
stellar pollution hypothesis in \S5. This test is also important to place a
more precise determination of the planetary radius and the mass of its core.
Finally, we summarize our results and discuss their implications in \S6.

\section{Computational method}
\label{sec:2}

We construct stellar models with the Eggleton stellar evolution code
\citep{Eggleton71, Eggleton72}. It is based on an 1-D Henyey scheme with
adaptive mesh. The input physics has been updated by \citet{Pols95} and
implemented by \citet{Sandquist02}. It includes combination of OPAL opacity for
stellar interior \citep{IR96} and low-temperature opacities for the surface
\citep{Alex94}, combination of SCVH \citep{Scvh95} and OPAL \citep{Rogers94}
equations of state, and nuclear reaction rates \citep{Bahcall95}. Chemical
diffusion is included, using the method of \citet{Thoul94}. Convection is
treated using the standard mixing-length theory and semi-convective mixing is
incorporated in a diffusive manner. By calibrating solar model with the
observational data of the present sun, we obtain the ratio of mixing length to
the pressure scale height $\alpha = 1.933$. We find satisfactory agreement when
comparing the radius of the convection zone base of our solar model with the
observed value from solar p-mode spectrum \citep{CD91}.

The code has been adapted to allow for the non-homogeneous metallicity for
polluted stellar models. In previous investigations on the effects of stellar
pollution (i.e. \citealt{laughlin97, Dotter03}; and \citealt{Cody05}), stellar
models are constructed based on the assumption that the high-metallicity
material accreted by the star is fully mixed in the stellar convective
envelope. This approximation is adequate for solar-type stars since all debris
are dissolved during their passage through the stellar outer convective layers.
HD\,149026 is an F-type star with a very shallow convective layer with a total
mass $\sim 0.003$\,$M_\odot$ in hydrogen gas. If the surface convective layer
is homogeneous and has an [Fe/H] as observed, its total mass of heavy elements
would amount to $\sim 10^{-4}$\,$M_\odot$, which is less than half of that
contained in HD\,149026b. If debris material plunges into the host star along
radial orbits, modest size planetesimals would survive the passage through the
convective layer and disintegrate in the radiative region. The amount of
disrupted planetesimals in the stellar atmosphere and envelope depends on their
trajectory, composition, as well as the background temperature and density
distributions. In order to take into account two limiting possibilities, we
adopt two simple prescriptions for the deposition of the accreted heavy
elements in the polluted stellar models.

We first consider the possibility that the residual planetesimals, embryos, and
ill-fated gas giants undergo a gradual orbital decay and enter the stellar
envelope through the stellar atmosphere with a large azimuthal motion. In this
case, even the largest embryos or gas giant will be completely disrupted in the
convective layer \citep{Sandquist02}. Thus in our first prescription, we assume
that the observed metallicity of HD\,149026 applies only to its thin convective
layer, whereas its interior has a solar metallicity.

In a second prescription, we consider the other possibility that pollutants
strike the host star on a direct radial trajectory. This assumption would be
appropriate if the debris is scattered into the stellar envelope by close
encounters with the gas giant. In this case, it is possible for a considerable
amount of pollutants to deposit in the star's vast radiative envelope below the
convective layer. For a swarm of planetesimals falling onto the star, we assume
a size distribution of $dN/da \sim a^{-3.5}$ \citep{Wetherill89}. This mass
distribution is comparable to those inferred for the asteroids and the Kuiper
belt objects and for the impactors of lunar craters \citep{Hartmann99}. We
assume that the residual planetesimals and embryos plunge radially towards the
center of the host star. Based on the results of previous simulations of
meteoritic impacts on the atmosphere of Venus \citep{Korycansky2005}, we assume
that the amount of ablated mass of individual planetesimals is comparable to
the mass of the background gas encountered along the path. Thus, the disruption
depth of planetesimals is a function of their sizes.

\begin{figure*}
\begin{center}
\epsscale{0.8}\plotone{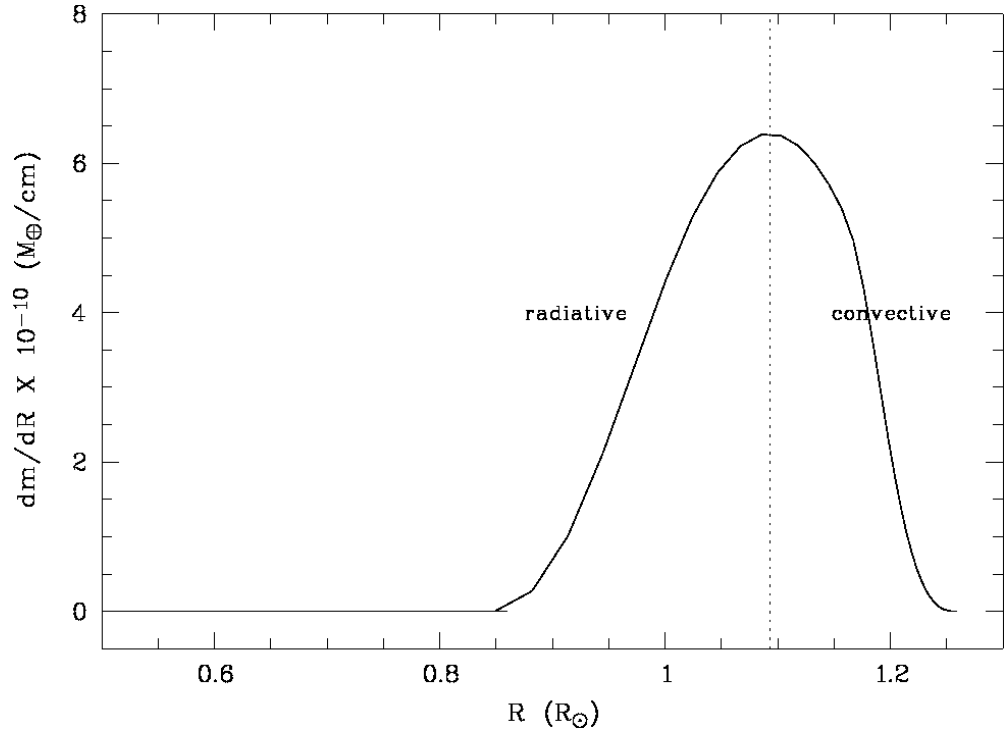}
\end{center}
\caption{Mass distribution of deposited planetesimals along stellar radius.
Totally 10\,$M_\oplus$ planetesimals were added to a 1.3\,$M_\odot$
star with an initial solar metallicity at ZAMS.}
\label{Fig:01}
\end{figure*}

Given the density distribution of a particular stellar structure, we can
integrate to obtain the deposited mass at each layer as a function of the
penetrating depth. Fig.~\ref{Fig:01} shows the mass distribution of $10
M_{\oplus}$ pollutants deposited along stellar radius for a $1.3 M_{\odot }$
star of an initial solar metallicity at ZAMS. The convective and radiative
zones are marked and separated by a vertical dotted line.

For both prescriptions, we consider the possibility that the bombardment of
debris may persist for an extended period of time. We assume the accretion
begins when the star reaches ZAMS with an initial solar and homogeneous
metallicity and the accretion lasts for a duration of 100 Myr. Observations in
the mid-IR have shown that the total mass of the grains in the disk declines
with time \citep{Rieke05}. We have thus assumed a time dependence of the rate
of planetesimal bombardment suggested by the IR observations. The accretion
rate thus declines monotonically as the star evolves.

\section{The effect of pollution on the stellar structure}
\label{sec:3}

For the first limiting case of the accreted planetesimals' trajectory, we mix
all the heavy elements of the accreted material through the convective zone.
In the second case, we still uniformly mixed those material which were ablated
in the convective region. While for the relatively large planetesimals
penetrating below the convective layer, we added the heavy elements deposited
in these regions to the local metallicity distribution in the radiative region.
Without a better understanding of the accreted material's composition, we
assume that it contains no hydrogen or helium but has the same composition of
heavy elements as that of the star in our calculation. The helium abundance of
the host star was deduced using ${\delta Y\over \delta Z} = {{Y-Y_{\odot}}\over
{Z-Z_{\odot}}} = 2.5$ \citep{Pagel92}.

\begin{figure}
\begin{center}
\epsscale{0.4}\plotone{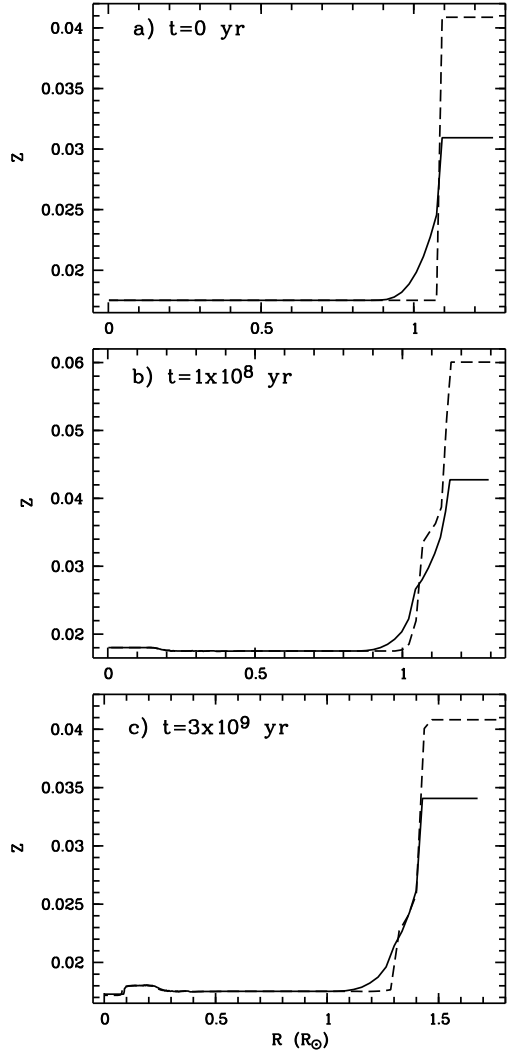}
\caption{Metallicity profiles against stellar radius at different
evolutionary times,
a) ZAMS, when $\sim 5 M_\oplus$ is accreted by the star,
b) $1\times 10^8 yr$, when totally 10\,$M_\oplus$ rocky material is added and
accretion stops thereafter,
c) $3\times 10^9 yr$.
Models polluted by two limiting ways are plotted.
The solid line shows the result of our second prescription for the distribution
of deposited material as plotted in Fig.~\ref{Fig:01}. While the dashed line
corresponds to our first prescription where all infalling material is dissolved
and mixed in the convective zone.}
\label{Fig:02}
\end{center}
\end{figure}

In Fig.~\ref{Fig:02}a), we show a comparison between the two limiting pollution
prescriptions of the initial metallicity profiles in the outer region of our
polluted stellar model. In each case, we assume $\sim 5 M_\oplus$ debris rocky
material is accreted onto a ZAMS host star. From Fig.~\ref{Fig:01} and
Fig.~\ref{Fig:02}a), we find that a considerable amount of the accreted
material enters the radiative region in the case that the planetesimals collide
the host star in a head-on trajectory. In the following one million year,
another 5\,$M_\oplus$ rocky material is added. Fig.~\ref{Fig:02} b) shows the
metallicity profiles of the two models at the time of $1\times 10^8 yr$, when
all 10\,$M_\oplus$ heavy elemental material has been accreted. At that time,
some diffusion has already occurred for both models at the boundary between the
convective and radiative regions. Fig.~\ref{Fig:02}c) shows the metallicity
profiles for both models at the time of $3\times 10^9 yr$, which is
approximately the age of HD\,149026. Due to the diffusion of heavy elements and
the deepening of the surface convective zone, the surface metallicity at this
late epoch has decreased considerably in both models.

\begin{figure}
\begin{center}
\epsscale{1}\plotone{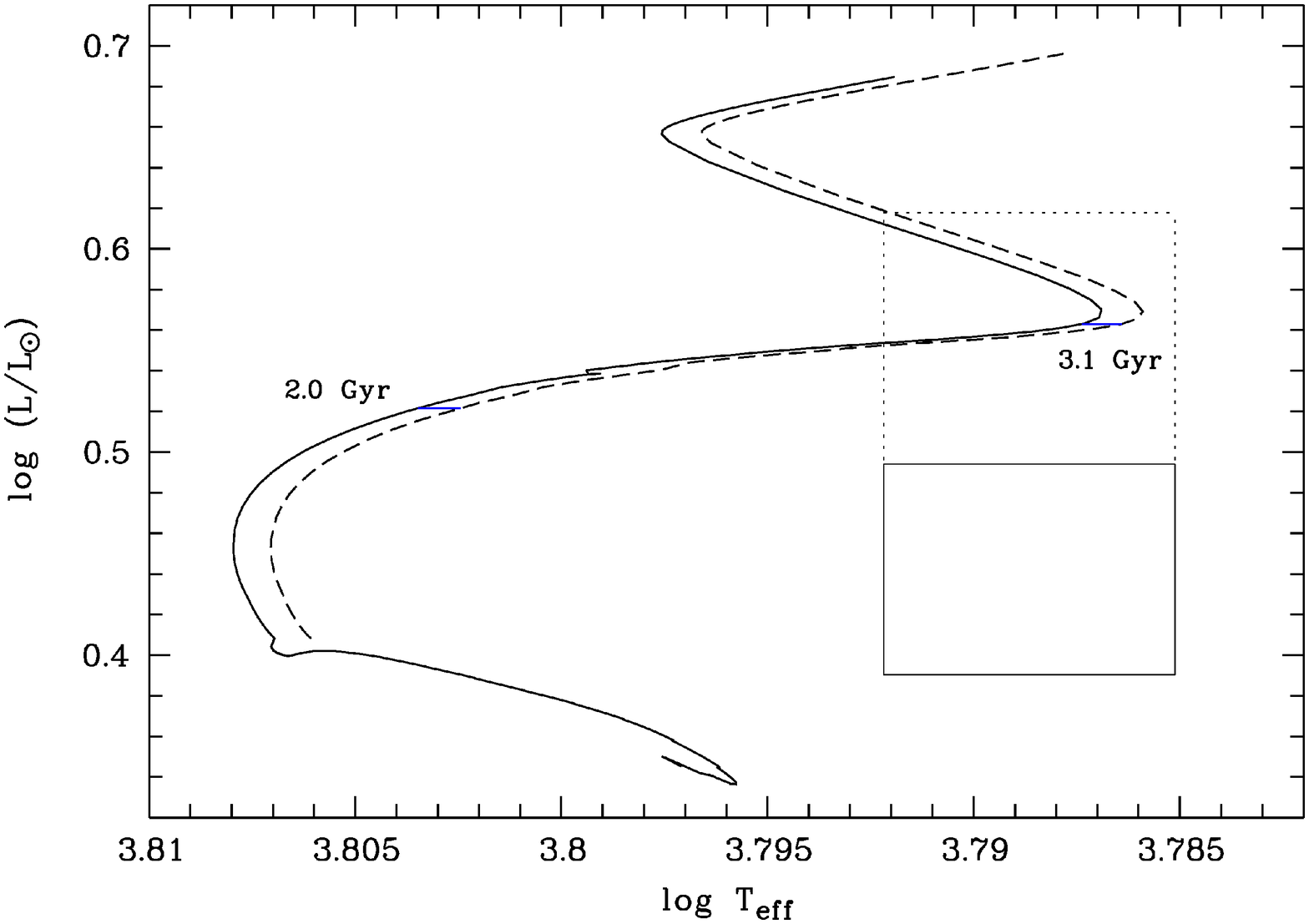}
\end{center}
\caption{Evolutionary tracks of a 1.3\,$M_\odot$ and solar interior 
metallicity star
polluted in two limiting ways. The total amount of accreted material
for both models is 10\,$M_\oplus$. Different line types have the same
denotations as that in Fig.~\ref{Fig:02}. The solid line
box shows 1$\sigma$ observational errors for luminosity and
temperature. The upper limit of luminosity taking into account
3$\sigma$ error of parallax is shown by the dotted box.}
\label{Fig:dif}
\end{figure}

Fig.~\ref{Fig:dif} shows the evolutionary tracks by adding the accreted
material in the two limiting ways. Both models have a mass of 1.3\,$M_\odot$
and an initial solar metallicity. We can find that during the main sequence,
especially where the tracks locate within the observational error bars for
HD\,149026, there is little difference between the two models in luminosity and
effective temperature. However, while the model with significant amount of
material deposited in the radiative zone can produce a surface metallicity
within the observed uncertainties for [Fe/H] of HD\,149026, the model with all
material mixed in the convective zone has a surface metallicity much larger
than the upper limit of the observed [Fe/H]. Therefore, the main effect of the
metallicity enhancement in the radiative zone compared with that in the
convection zone is that it can increase the radiative opacity, redden the star
and make it expand without causing significant surface metallicity
increasement. As HD\,149026 has a very shallow surface convective zone, it's
more likely that a lot of infalling planetesimals will penetrate into the
radiative region. Therefor, we use the second pollution prescription for the
following calculations.

\section{Evolution of HD\,149026 \label{sec:4}}

Based on the above definitions of pollution, we construct polluted stellar
models for HD\,149026 which host a Saturn-mass planet with a large dense core
in a 2.8766 day orbit \citep{Sato05}. Three basic observational parameters are
required to constrain our models. From spectroscopic analysis \citep{Sato05},
the effective temperature $T_{eff}$ and metallicity [Fe/H] for HD\,149026 are
determined to be $6147 \pm 50$ K and $0.36 \pm 0.05$, respectively. Another
important parameter is its luminosity. From the Hipparcos catalog
\citep{ESA97}, the parallax and visual magnitude of HD\,149026 are $12.68\pm
0.79$ mas and 8.15, respectively. The bolometric correction (BC) is estimated
to be $-0.026\pm 0.006$ by interpolating the effective temperature using the
$T_{eff}-BC$ conversion relationship provided by \citet{Flower96}. Then, we
arrived at a bolometric luminosity of $2.756^{+0.363}_{-0.300} L_\odot$ with
1$\sigma$ error bar of the parallax. We show below the results of a range of
theoretical models with different stellar parameter (such as stellar mass and
interior metallicity) that are within the observational uncertainties. For
example, a 3$\sigma$ uncertainty in the parallax-distance determination leads
to an uncertainty of $(+1.393,-0.789) L_\odot$ in the absolute luminosity of
HD\,149026.

For the standard model of HD\,149026 without pollution (with a homogeneous
metallicity distribution), we obtain a stellar mass of 1.3\,$M_{\odot}$ and an
age of 2.27 Gyr. Around this standard model, we have constructed a series of
polluted stellar evolution models with masses in the range of 1.2 to
1.4\,$M_{\odot}$. Two sets of initial metallicity, solar metallicity and 1.5
times solar metallicity were specified for these models. The models were
evolved by accreting different amount of heavy elements.

\begin{figure}
\begin{center}
\epsscale{1}\plotone{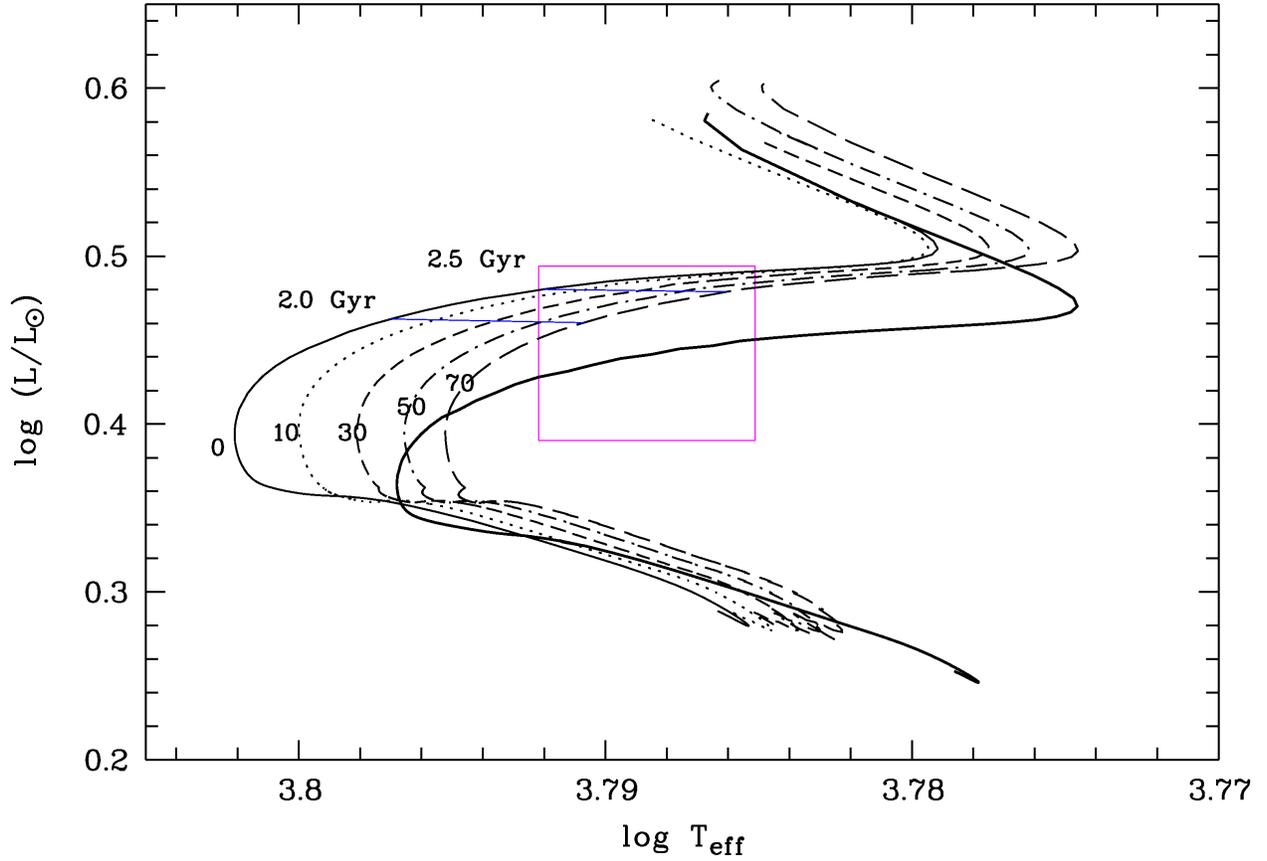}
\end{center}
\caption{The unpolluted models with 1.3\,$M_\odot$ stellar mass and
internal 1.5 solar
metallicity are plotted in the thin solid line. The tracks marked by different
numbers denote different amount of pollution in earth mass based on the
unpolluted model. While the model with the mass of 1.3\,$M_\odot$ and observed
metallicity of HD\,149026
is plotted in thick line for comparison. Overplotted is the 1$\sigma$
error box for observed effective temperature and luminosity.}
\label{Fig:04}
\end{figure}

We firstly illustrate the effect of pollution of the stellar-evolution track in
the color-magnitude diagram. In Fig.~\ref{Fig:04}, we show a series of models
with $M_\ast=1.3 M_{\odot}$ and $Z_\ast=1.5 Z_{\odot}$ throughout the star. In
this figure, the evolutionary track of a model with an unpolluted homogeneous
interior (with 1.5 solar metallicity) is denoted by a thin solid line.
Overplotted are the evolutionary tracks of models which are polluted by a total
amount 10\,$M_{\oplus}$, 30\,$M_{\oplus}$, 50\,$M_{\oplus}$ and
70\,$M_{\oplus}$ of heavy elements. Two ages are labeled on the tracks. For
comparison, we have also computed the evolutionary track of a 1.3\,$M_\odot$
model with the observed metallicity of HD\,149026 (the thick solid line in Fig.
\ref{Fig:04}).

Contrasting these models at any given effective temperature, we find that the
accretion of heavy elements decreases the luminosity as a consequence of the
opacity enhancement in the radiative layer. The reduced efficiency of radiation
transfer also increases the central temperature of the star and speeds up the
stellar evolution process. Consequently, at the same age, the surface
metallicity enhancement has the general effects of 1) reducing the effective
temperature near the main sequence turn off; 2) increasing the stellar radius;
and 3) decreasing the luminosity.

\section{Observational tests}
\label{sec:5}

The accretion of heavy elements increases the metallicity of the surface layer
while that in the deep stellar interior retains its initial value. In order to
make direct comparison with the observed properties of HD\,149026, we need to
match the surface metallicity of a series of models with the observed value
[Fe/H]=0.36. This surface boundary condition places a constraint on the extent
of heavy-element pollution. Take the 1.3\,$M_{\odot} $ models with internal 1.5
$Z_\odot$ for example, the amount of pollution needed to satisfy all the
observational constraints ($L_\ast$, $T_{\rm eff}$, and [Fe/H]) within
1$\sigma$ error bars is between 10 to 20\,$M_{\oplus}$. 

\begin{figure}
\begin{center}
\epsscale{1}\plotone{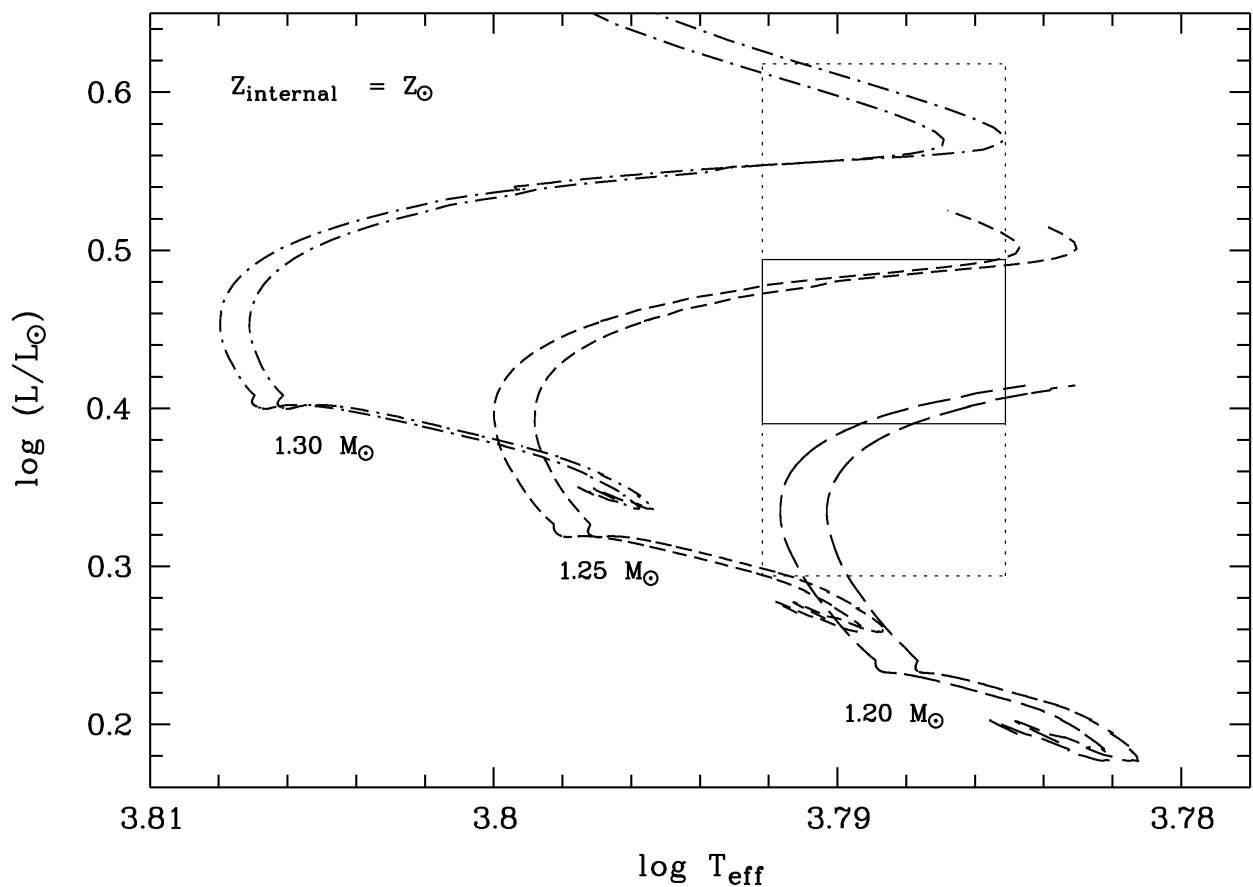}
\end{center}
\caption{Possible polluted models with internal solar metallicity for different
stellar masses. For each mass, the two tracks correspond to the
upper and lower limit of metallicity that needs be added to match the
observed range in [Fe/H].
The 1$\sigma$ error box for observed effective temperature and
luminosity is plotted in solid line. While the dotted error box takes into 
account 3$\sigma $ error of parallax.}
\label{Fig:05}
\end{figure}

We note that the relative errors in both [Fe/H] and $L_\ast$ are an order of
magnitude larger than that in $T_{\rm eff}$. We firstly consider the possible
range of stellar models. In Fig.~\ref{Fig:05}, We consider a series of models
(with different $M_\ast$) with initial solar metallicity which is also
preserved in the stellar interior. For each mass, the two tracks correspond to
the upper and lower limit of metal required to match the observed range in
[Fe/H]. These nearly overlapping tracks indicate that the values of $L_\ast$
and $T_{\rm eff}$ are more sensitive to $M_\ast$ than to the exact value of
[Fe/H]. This plot also indicates that the amount of pollution can be
constrained quite well by the observed metallicity for a model with the same
mass and internal metallicity.

Overplotted in Fig.~\ref{Fig:05} is the 1$\sigma$ (solid lines) error box for
the observed effective temperature and luminosity. Considering that the stellar
distance is difficult to be determined accurately, we extended the error box by
taking into account 3$\sigma $ error of parallax (dotted lines). Models with
$M_\ast > 1.35$\,$M_\odot$ are excluded because they produce over luminous (for
the observed range of $T_{\rm eff}$) tracks. Models with $M_\ast <
1.15$\,$M_\odot$ are excluded because they produce insufficiently hot $T_{\rm
eff}$ (for the observed range of $L_\ast$) tracks. A star of
$M_\ast=1.22$\,$M_\odot$ passes the center of the error bar box. For this
particular model, the total amount of pollutant required to bring the surface
metallicity to the observed value is $\Delta M_Z \simeq 28 M_\oplus$, which is
a significant fraction of the heavy element content in HD\,149026b.

\begin{figure}
\begin{center}
\epsscale{1}\plotone{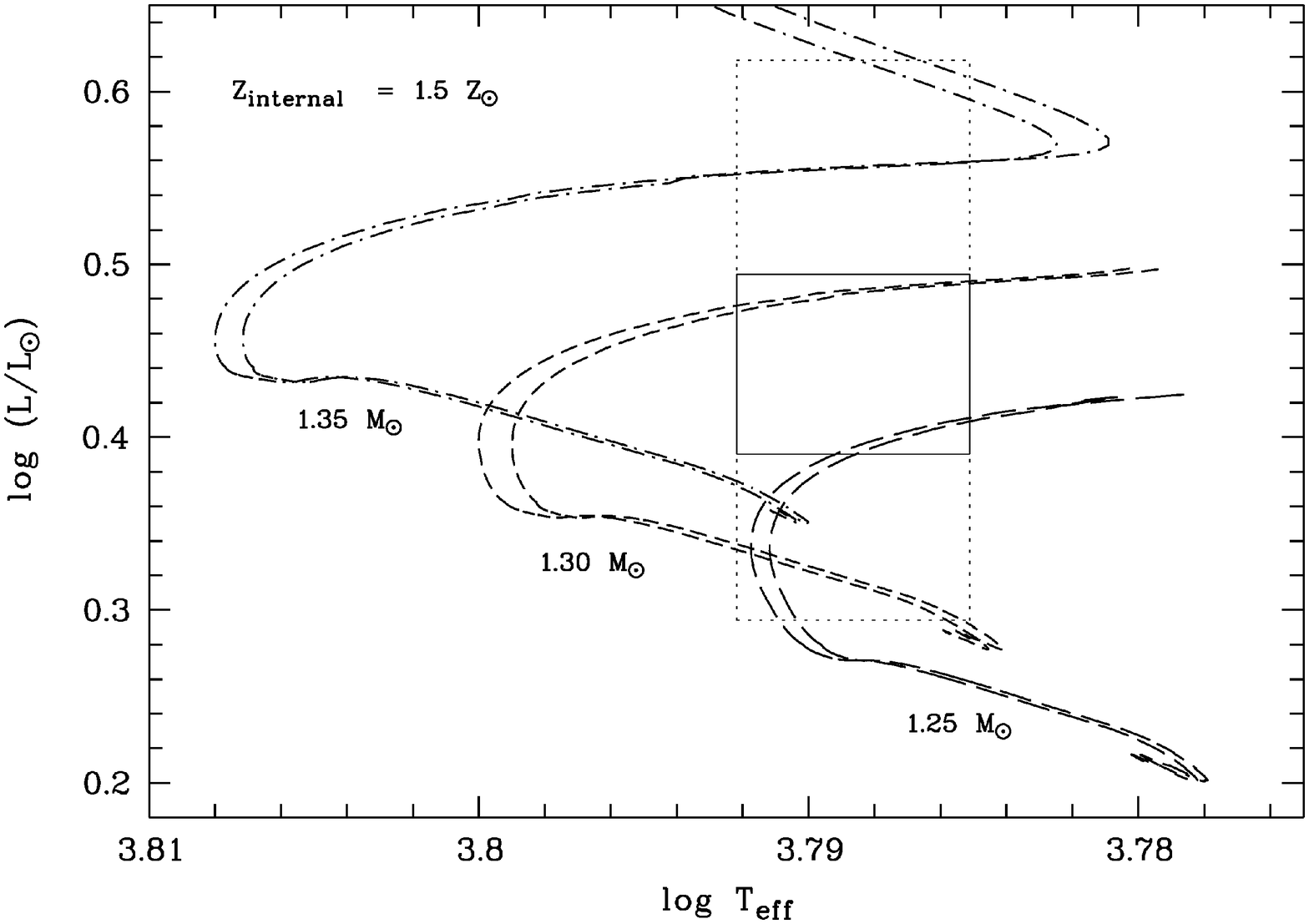}
\end{center}
\caption{Same as Fig.~\ref{Fig:05}, but for possible polluted models with 
an internal metallicity of 1.5 times solar.}
\label{Fig:06}
\end{figure}

We considered another series of models with an initial metallicity equals to
1.5 times that of the Sun (Fig.~\ref{Fig:06}). Obviously, less heavy elements
are needed to match the observed [Fe/H] than those models in Fig.~\ref{Fig:05}.
Models with $1.2 M_\odot < M_\ast < 1.4 M_\odot$ pass through the extended
error box with $M_\ast = 1.27 M_\odot$ as a best fit, which requires a total
amount of pollutant of $\Delta M_Z \simeq 20 M_\oplus$. Note that the
unpolluted model in Fig.~\ref{Fig:04} which matches the most probable
observational data has $M_\ast = 1.3 M_\odot$.

The above results indicate that several models, with or without pollution, can
be constructed within the present observational uncertainties. While effective
temperature and surface metallicity have been determined quite well by present
spectroscopic observations, the current observational uncertainties in
luminosity make it difficult to distinguish the models. More accurate
luminosity is thus highly desired. We note from Figs.~\ref{Fig:05} and
\ref{Fig:06} that stellar mass and internal metallicity are degenerate
parameters in determining the evolutionary tracks in H-R diagram, even if
stellar luminosity is well constrained. Given independently measured stellar
mass with high precision, the internal metallicity of the star can be
constrained by fitting the theoretical evolutionary tracks. A definite
difference between the internal and surface metallicity in a star is a good
evidence of the occurrence of pollution onto this star. Thus, a set of more
accurately determined luminosity and independently measured stellar mass,
combined with our stellar models, provide a promising way to explore the
pollution (if any) history of the star.

The stellar luminosity adopted in our models are calculated using the parallax
and visual magnitude obtained from the Hipparcos catalog \citep{ESA97}. For
HD\,149026 with a visual magnitude of 8.15, the uncertainty of parallax is $\pm
0.79$ mas. Higher accuracy will be achieved by the European Gaia astrometry
mission which is due for launch in 2011. For stars with $V=10$, the parallax
accuracy is expected to be around 7 $\mu$as, which is less than 1\% of present
observational uncertainty \citep{Jordi2006}. This improvement reduces the
uncertainty in luminosity by a factor of 30. In this case, if the stellar mass
can be determined independently with an accuracy of $\sim 1\%$, it's possible
that the polluted and unpolluted models are distinguishable.

However, it is a big challenge to measure stellar mass precisely. Usually,
stellar mass can be determined in binary systems using astrometric methods. For
eclipsing binaries, an accuracy of 2-3\% can be achieved in the best case
\citep{Balega2007}. For single stars with a planetary companion, it is possible
to determine the stellar mass if the planet transits the star. By solving the
equations of transit geometry and Kepler's third law, the average stellar
density can be derived as \begin{equation} \rho_\ast = {32 P \over \pi
G}{{\Delta F ^{3/4}} \over {(t_T^2 - t_F^2)^{3/2}}} \end{equation} where $t_T$
is the total transit duration, $t_F$ is the duration of the transit completely
inside ingress and egress, $P$ is the period, and $\Delta F$ is the transit
depth \citep{Seager03, Gillon06, Winn07}. The stellar radius can be determined
independently by using the Stefan-Boltzmann law, rather than the isochrone
fitting method \citep{Sato05}. Combined with $\rho_\ast$ calculated from the
transit light curve, the stellar mass $M_\ast$ can be obtained independently.
At present, the stellar density can be constrained to 3\% by HST observation
\citep{Pont2007}. Also, the stellar radius can be constrained better by the
improved luminosity measurement. Thus, it's possible that the stellar mass can
be obtained with a marginal accuracy as required to test our theoretical
models.

Gravitational microlensing is another promising method to determine the stellar
mass of single stars accurately. \citet{Paczynski1998} shows that it's possible
for the Space Interferometry Mission (SIM) to measure $M_\ast $ of nearby stars
with an accuracy of 1\% using this method.

\begin{figure}
\begin{center}
\epsscale{1}\plotone{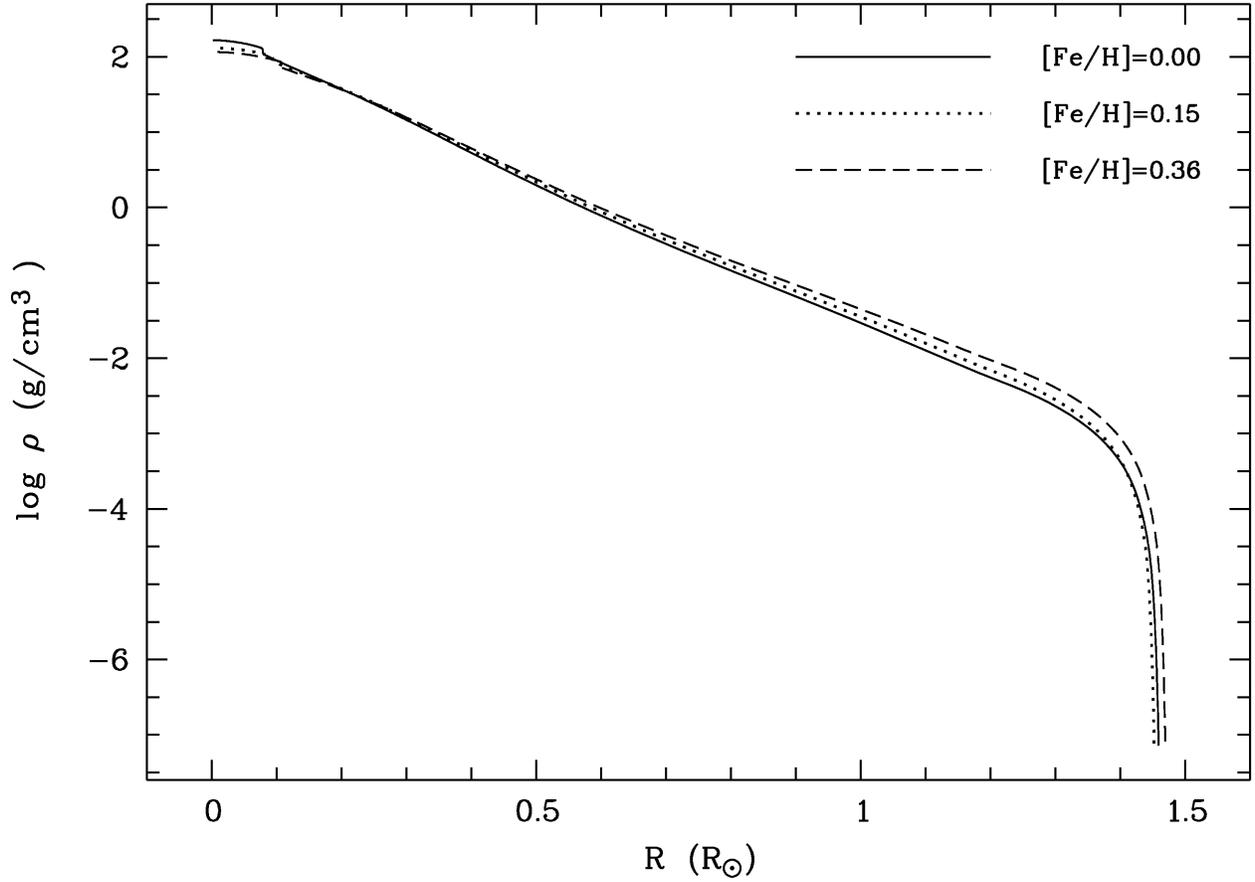}
\end{center}
\caption{Density profiles for the three best-fit models, with stellar mass
$M_\ast = 1.22 M_{\odot }$ and initial [Fe/H] = 0 (solid line), 
$M_\ast = 1.27 M_{\odot }$ and [Fe/H] = 0.15
(dotted line), and $M_\ast = 1.3 M_{\odot }$ and 
[Fe/H] = 0.36 (dashed line), respectively.}
\label{Fig:rho}
\end{figure}

\begin{figure}
\begin{center}
\epsscale{1}\plotone{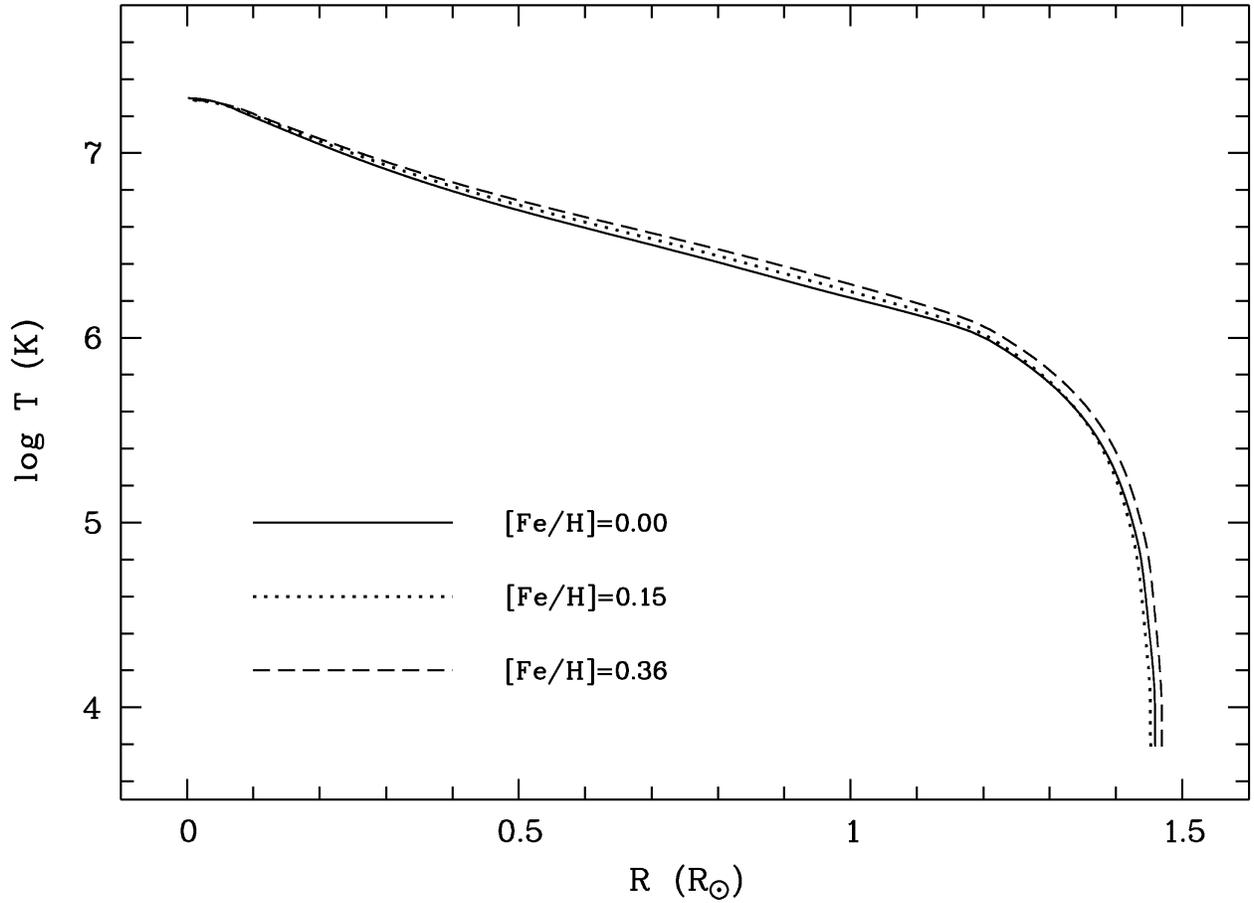}
\end{center}
\caption{Temperature profiles for the three best-fit models, 
with stellar mass
$M_\ast = 1.22 M_{\odot }$ and initial [Fe/H] = 0 (solid line),
$M_\ast = 1.27 M_{\odot }$ and [Fe/H] = 0.15
(dotted line), and $M_\ast = 1.3 M_{\odot }$ and
[Fe/H] = 0.36 (dashed line), respectively.}
\label{Fig:t}
\end{figure}

With the aforementioned upcoming high-precision astrometry missions, we expect
that our theoretical stellar models will be able to test whether pollution is a
significant effect inducing the high metallicities of stars with planets.
Another direct probe for testing the occurrence of pollution is associated with
the modification in the internal structures and the oscillation frequencies of
the stars. In Figs.~\ref{Fig:rho} and \ref{Fig:t}, we plot the density and
temperature distribution in the stellar envelope for three best-fit models
(with initial and internal [Fe/H] = 0, 0.15, and 0.36, and stellar mass $M_\ast
= 1.22 M_{\odot }$, $1.27 M_{\odot }$, and $1.3 M_{\odot }$ respectively). We
will consider elsewhere the difference in the g-mode oscillation frequency
associated with the dispersion in the structure of these models.

\section{Summary and Discussions}
\label{sec:6}

In this paper, we considered the possibility that the outer envelope of a
metal-rich planet-bearing star HD\,149026 may have been significantly polluted
by residual planetesimals or some gas giant planets. This system is
particularly interesting because it contains a planet with most of its mass in
its core and an intermediate-mass star which has recently entered the post main
sequence evolutionary phase. In a subsequent paper, we will suggest that this
large-core Saturn-mass planet HD\,149026b was formed through giant impacts onto
a proto gas giant planet by either residual protoplanetary embryos or by other
gas giant planets in the proximity of the host star. A significant fraction of
metal of the building blocks may not merge or be retained as a consequence of
grazing high-velocity collisions. However, those debris may undergo orbital
decay or remain in the close proximity of HD\,149026b. 

We assume that a substantial fraction of the original residual planetesimals in
the neighborhood of HD\,149026b may have been scattered into its host star
which is an F stars with a very shallow convective layer. We carried out
evolution calculations to take into account the effect of stellar pollution.
We suggest that in this and other F star with planets, the extent of stellar
pollution by residual planetesimals may be tested by accurate determinations of
their mass, luminosity, and effective temperature. Such accurate measurements,
when combined with an accurate transit light curve, may provide a firm support
on the stellar accretion of a substantial population of
terrestrial-planet-building material as well as one or more gas giant planets
similar to HD 149026b.

\acknowledgements We thank Drs C. Agnor, P. Bodenheimer, L.~C. Deng,
D. Korycansky, and E. Sandquist for useful conversation. This work is
supported by NASA (NAGS5-11779, NNG04G-191G, NNG06-GH45G, NNX07AL13G,
HST-AR-11267), JPL (1270927), and NSF(AST-0507424).

%%%%%%%%%%%%%%%%%%%%%%

\end{document}